\documentclass[aps,twocolumn,superscriptaddress,amsmath,amsfonts,
               floatfix]{revtex4}

\usepackage{epsfig,psfrag}
\usepackage{CJK}

\usepackage{graphicx}
\usepackage{dcolumn}
\usepackage{bm}
\usepackage{amssymb}

\begin{document}
\title{An explanation for high defect tolerance in metal halide perovskite quantum dots}
\author{Yu Cui}

\author{Xiao-Yi Liu}

\author{Jia-Pei Deng}

\author{Zhi-Qing Li}

\author{Zi-Wu Wang*}
\affiliation{Tianjin Key Laboratory of Low Dimensional Materials Physics and Preparing Technology, Department of Physics, School of Science, Tianjin University, Tianjin 300354, China}
\email{wangziwu@tju.edu.cn}

\begin{abstract}
 We propose Auger-like process assisted by quantum defects in metal halide perovskite quantum dots, where a charge carrier in the ground state of the quantum dot is trapped by quantum defects, resulting in another charge carrier in defect is excited and returns back to the ground state of the quantum dot. We find that the whole process is on the femtosecond scale. More importantly, the process is independent of the depth and species of the defects, which is in good agreement with the recent theoretical prediction using \emph{ ab initio} nonadiabatic molecular dynamics simulation. This Auger-like process may provide a potential explanation of high defect tolerance in metal halide perovskite materials.
\end{abstract}

\maketitle
\section{Introduction}
As the ideal candidates for many potential applications on optoelectronic and photovoltaic devices, metal halide perovskite quantum dots (MHPQDs) have spurred intense research efforts arising from their extraordinary properties, such as the bright photoluminescence covering the entire visible spectral range, strong optical absorption coefficient, and ease of fabrication\cite{wx1,wx2,wx3,wx5,wx8}. In particular, suffering from varieties of defects that are inevitable during the typical growth processes, these outstanding properties are still preserved. Meanwhile, this exceptional defect tolerance plays a key role in high photoluminescence quantum yield\cite{wx6,wx9,wx10,wx11}. However, the underlying physics for this high defect tolerance (HDT) is ambiguous.

In fact, several mechanisms have been proposed, attempting to give reasonable explanations for this HDT in metal halide perovskites materials in the past years. Tan $et$ $al$. attributed it to the influence of dipolar cation\cite{wx11}. They found that the dipolar cation reoriented in response to the local electrical field on account of the electrostatic interaction between the dipolar cation and the local electrical field generated by the defects, which reduced the capture cross-section of nonradiative recombination. From the electronic structure point of view, metal halide perovskites materials are lack of the bonding-antibonding interactions between the conduction and valence bands, resulting in most defect's levels are within the conduction or valence bands, and thus maintain the clean bandgaps\cite{wx12,wx28}.  With the strong evidence of the large polaron by different experimental techniques in metal halide perovskites materials\cite{wx13,jhz1,jhz2,jhz3,jhz4,jhz5}, the HDT also has been widely proposed to relate to the formation of large polaron, which reduces the scattering of charge carriers from defects. Recently, Chu $et$ $al$ pointed out that the HDT may stem from the photogenerated carriers only coupled with low-frequency phonon modes, giving rise to the notably decreasing of the nonadiabatic coupling between the donor and acceptor states, so both the pristine and defective systems have a long electron-hole recombination time\cite{wx14}. Although the HDT has been speculated quantitatively by several mechanisms above mentioned, the numerical evaluations for the trapping lifetime of charge carriers by defects are still very few.
\begin{figure}[htbp]
	\centering
	\includegraphics[width=2.8in,keepaspectratio]{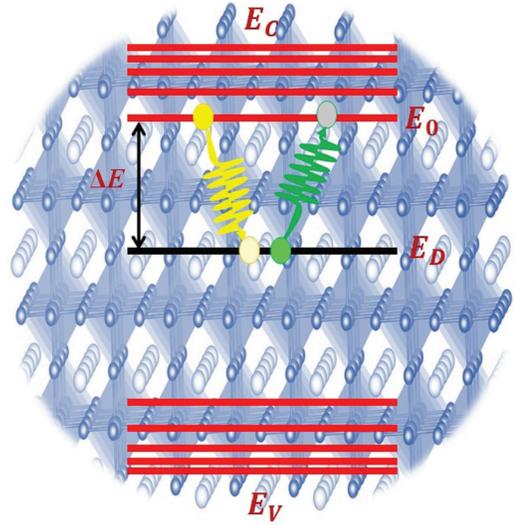}
	\caption{\label{compare} The schematic diagram of Auger-like recombination process assisted by the defect in a metal halide perovskite quantum dot. $E_C$ and $E_V$ denote the conduct and valence bands, respectively. $E_0$ is the ground state energy level of the quantum dot, $E_D$ is the defect level and $\Delta E=E_0-E_D$ denotes the depth of defect.}
\end{figure}

In the present paper, we propose the Auger-like nonradiative processes mediated by the neutral, positive, and negative defects in MHPQDs, respectively. We give the dependences of the lifetime of Auger-like process on the depth of defect in the bandgap and types of the defects. We also discuss the influence of the radius of quantum dot on the lifetime. These theoretical results not only provide the good explanation to defect tolerance in MHPQDs, but also give some insight for modulating the performances of MHPQDs-based devices.

\section{theoretical model}
As schemed in Fig. 1, the whole process of the Auger-like recombination mediated by the quantum defect can be divided into two steps: (1) a charge carrier in the ground state of the MHPQDs is trapped by quantum defects; (2) another charge carrier in defect is excited by absorbing the energy produced in the trapping process, and then falls back into the ground state of the quantum dot. According to the classical Auger process\cite{wx15,wx16,wx17}, this Auger-like process can be expressed as
 \begin{equation}
 {\tau ^{{\rm{ - 1}}}}{\rm{ = }}\frac{{{\rm{2}}\pi }}{\hbar }{\left| {{M_{Direct}} - {M_{Exch}}} \right|^2}\delta ({E_D} - {E_0}),
 \end{equation}
 with
 \begin{eqnarray}
{M_{Direct}} &=& \frac{{{e^2}}}{{4\pi {\varepsilon _0}{\varepsilon _r}}}\int {\int {\Psi _0^*({{\bf{r}}_{\bf{1}}})} } \Phi _0^*({{\bf{r}}_{\bf{2}}})\frac{1}{{\left| {{{\bf{r}}_{\bf{1}}}{\bf{ - }}{{\bf{r}}_{\bf{2}}}} \right|}}\nonumber\\
&&\times\Psi _0({{\bf{r}}_{\bf{1}}})\Phi _0({{\bf{r}}_{\bf{2}}})\rm{d}{{\bf{r}}_{\bf{1}}}\rm{d}{{\bf{r}}_2},
 \end{eqnarray}
 and
 \begin{eqnarray}
 {M_{Exch}} &=&  \frac{{{e^2}}}{{4\pi {\varepsilon _0}{\varepsilon _r}}}\int {\int {\Psi _0^*({{\bf{r}}_{\bf{1}}})} } \Phi _0^*({{\bf{r}}_{\bf{2}}})\frac{1}{{\left| {{{\bf{r}}_{\bf{1}}}{\bf{ - }}{{\bf{r}}_{\bf{2}}}} \right|}}\nonumber\\
 &&\times\Psi _0({{\bf{r}}_2})\Phi _0({{\bf{r}}_1})\rm{d}{{\bf{r}}_{\bf{1}}}\rm{d}{{\bf{r}}_2},
 \end{eqnarray}
where $\tau$ is the lifetime of the whole process, ${M_{Direct}}$ and ${M_{Exch}}$ represent the direct and  exchange terms for the electron-electron Coulomb interaction between the initial state and the final state, respectively. ${{\bf{r}}_{\bf{1}}}$ and ${{\bf{r}}_{\bf{2}}}$ are position variables, $E_D$ is the defect level and $E_0$ is the ground state energy level in a quantum dot, the difference between them $\Delta E=E_D-E_0$ denotes the depth of the defect. $e$ is the charge carrier, ${\varepsilon _0}$ is the permittivity of vacuum and ${\varepsilon _r}$ is the relative permittivity. The ground-state wave function ${\Phi _0}$ in MHPQDs is given by\cite{pqd1,pqd2} (the detail processes are shown in supplemental materials)
\begin{equation}
{\Phi _0}(r) = \frac{1}{{{\pi ^{3/4}}R_0^{3/2}}}\exp ( - \frac{{{r^2}}}{{2R_0^2}}),
\end{equation}
where $R_0$ is the radius of quantum dot.

Following the quantum defect model\cite{qdm1,qdm2,qdm3}, the ground-state wave function ${\Psi _0}$ for defect with arbitrary binding energy can be expressed as
\begin{equation}
{\Psi _0}(r)=\mathcal{N}(\frac{r}{a^*})^{\mu-1}\exp({-\frac{r}{\vartheta a^*}}),
\end{equation}
with
\begin{equation}
\mathcal{N}=\frac{1}{\sqrt{4\pi a^{*3}(\vartheta/2)^{2\mu+1}\Gamma(2\mu+1)}}\nonumber\\,
\end{equation}
\begin{equation}
a^*=\frac{4\pi\varepsilon\hbar^2}{e^2m^*}\nonumber\\,
\end{equation}
\begin{equation}
\vartheta  = \frac{{{e}}}{{\sqrt {8\pi \varepsilon{a^*}\Delta E} }}\nonumber\\,
\end{equation}
where $N$, $a^*$, and $\vartheta$ represent the normalization constant, effective Bohr radius, and quantum defect parameter, respectively. $m^*$ is the effective mass of electron (or hole), types of defect are reflected by the parameter $\mu$ = +$\vartheta$ for the positive defect, $\mu$ = -$\vartheta$ for the negative defect, and $\mu$ = 0 for the neutral defect.

To calculate the ${M_{Direct}}$ and ${M_{Exch}}$ directly, the term $1/\left| {{{\bf{r}}_{\bf{1}}}{\bf{ - }}{{\bf{r}}_{\bf{2}}}} \right|$ is expressed in Fourier series \cite{wx18,wx19}
\begin{equation}
\frac{1}{{\left| {{{\bf{r}}_{\bf{1}}} - {{\bf{r}}_{\bf{2}}}} \right|}} = \frac{1}{{{{(2\pi )}^3}}}\int {\frac{{4\pi }}{{{q^2}}}} \exp [i{\bf{q}} \cdot {\bf{(}}{{\bf{r}}_{\bf{1}}}{\bf{ - }}{{\bf{r}}_{\bf{2}}}{\bf{)}}]{\rm{d}}{\bf{q}},
\end{equation}
where ${\bf{q}}$ represents the electronic wave vector. Substituting Eqs. (4)-(6) into Eqs. (2) and (3) leads to
\begin{widetext}
\begin{eqnarray}
{M_{Direct}} = \frac{{8{N^2}{e^2}}}{{{\pi ^{3/2}}R_0^3{\varepsilon _0}{\varepsilon _r}}}\int {\int {\int {{{\left( {\frac{{{r_2}}}{{{a^*}}}} \right)}^{2\mu  - 2}}\exp \left( { - \frac{{r_1^2}}{{R_0^2}} - \frac{{2{r_2}}}{{\vartheta {a^*}}}} \right)\frac{{\sin \left( {q{r_1}} \right)\sin \left( {q{r_2}} \right)}}{{{q^2}{r_1}{r_2}}}r_1^2r_2^2d{r_1}d{r_2}dq} } } ,
\end{eqnarray}
for the direct term and
\begin{eqnarray}
{M_{Exch}} = \frac{{8{N^2}{e^2}}}{{{\pi ^{3/2}}R_0^3{\varepsilon _0}{\varepsilon _r}}}\int {\int {\int {{{\left( {\frac{{{r_1}{r_2}}}{{{a^{*2}}}}} \right)}^{\mu  - 1}}\exp \left( { - \frac{{{r_1} + {r_2}}}{{\vartheta {a^*}}} - \frac{{r_1^2 + r_2^2}}{{2R_0^2}}} \right)\frac{{\sin \left( {q{r_1}} \right)\sin \left( {q{r_2}} \right)}}{{{q^2}{r_1}{r_2}}}} } } r_1^2r_2^2d{r_1}d{r_2}dq,
\end{eqnarray}
\end{widetext}
for the exchange term, respectively.

In this paper, we choose typical CH$_3$NH$_3$PbI$_3$ (MAPbI$_3$) and CH$_3$NH$_3$PbBr$_3$ (MAPbBr$_3$) quantum dots as examples to calculate the lifetime of this Auger-like process. The adopted values of parameters for two materials are shown in Table ${\rm I}$.

\begin{table}[!b]
\caption{\label{compare} The adopted parameters for the theoretical calculation of the Coulomb matrix element. ${\varepsilon _0} = 8.85 \times {10^{ - 12}}$ F/m is the permittivity of free space and $m_0 = 9.1 \times {10^{-31}}$ kg is the free electron mass. Parameters were taken from Refs. \cite{wx21} and \cite{wx22}.
	}
\setlength{\tabcolsep}{2mm}{
	\begin{tabular}{ccccccccc}
	
		\hline
		\hline
		Parameter                  & MAPbI$_3$     &  MAPbBr$_3$     \\[0.8ex]\hline
		Effective mass ($m^*$)     &   0.2$m_0$\cite{wx21}  &   0.23$m_0$\cite{wx21}   \\
		Permittivity ($\varepsilon$)     &   33.5${\varepsilon _0}$\cite{wx22}  &   32.3${\varepsilon _0}$\cite{wx22}  \\
		Effective Bohr radius ($a^*$)       & 8.9 nm\cite{wx22} & 7.5 nm\cite{wx22} \\
		Quantum defect parameter ($\vartheta $) & Variable & Variable \\
		Quantum dots radius ($R_0$)               & Variable & Variable\\
		\hline
		\hline
	\end{tabular}}
\end{table}

\begin{figure*}[!t]
	\includegraphics[width=6.5in,keepaspectratio]{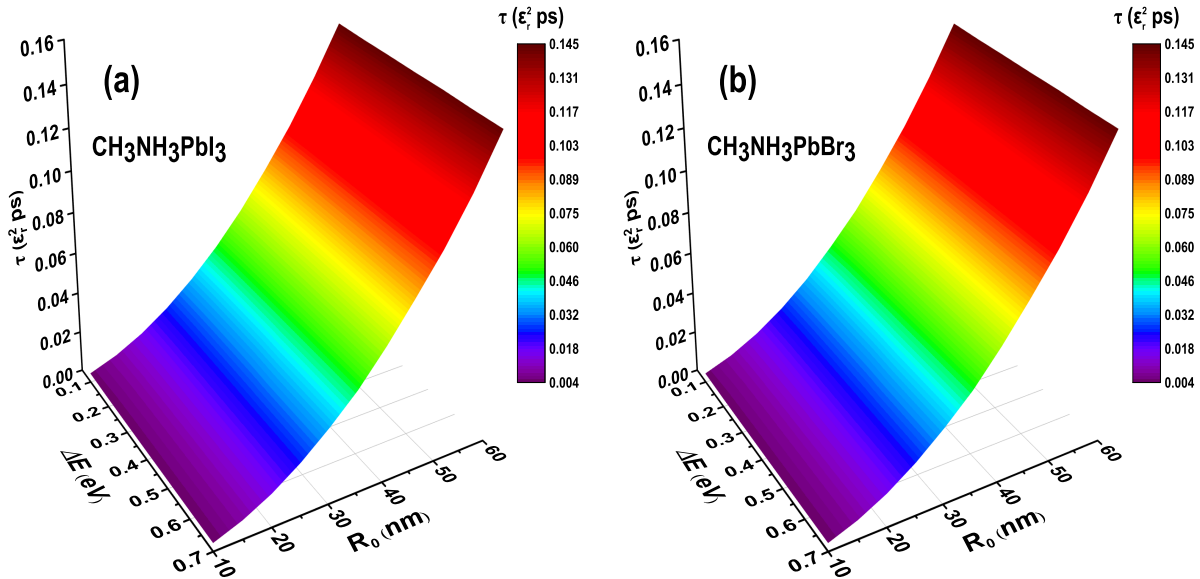}
	\caption{\label{compare} The lifetime of Auger-like processes $\tau$ as functions of the depth of the positive defect $\Delta E$ and the quantum dot radius $R_0$ in CH$_3$NH$_3$PbI$_3$ (a) and CH$_3$NH$_3$PbBr$_3$ (b), respectively.}
\end{figure*}

\section{Results and discussion}
Figs. 2 (a) and (b) depict the lifetime $\tau$ as functions of the depth of positive defect levels $\Delta E$ and the radius of quantum dot $R_0$ in MAPbI$_3$ and MAPbBr$_3$, respectively. One can find that the whole Auger-like process, including the trapping and detrapping processes of a charge carrier, is very fast on the femtosecond time scale. This ultrafast process results in an illusion that the density of charge carriers in the conduction band (or valence band) is unchanged, which seems that defects have no influence on the charge carrier densities. Here, we must emphasize that the lifetime depends on the permittivity ${\varepsilon _r}$ of the electron-electron Coulomb interaction, which can vary in the range of 5 $\thicksim$ 30 proved by several experiments\cite{wx22,wx23,wx24}, depending on the dielectric environment sensitively\cite{wx26,wx27}. However, the lifetime ($\varepsilon_r^2\tau$) still keeps on picosecond time scale even at $\varepsilon_r=30$ in our numerical calculations. More importantly, the lifetime is independent of the depth of the defects in the bandgap as shown in Fig. 2 for two materials. The similar results are obtained for the negative and neutral defects shown in the supplemental materials. This indicates that the lifetime does not change regardless of the different types of defects introduce a shallow or deep state, suggesting the ``high'' property of defect tolerance accurately again. Recently, Chu $et$ $al$\cite{wx14} also showed that charge recombination does not depend on the types of defects and their locations in the band gap based on \emph{ ab initio} nonadiabatic molecular dynamics simulation. They attributed this defect tolerance to the nature of the soft inorganic lattice of these metal halide perovskites with small bulk modulus, which gives rise to the photogenerated carriers are inclined to couple with low-frequency phonons. Hence, our theoretical results provide another explanation for the high defect tolerance and enrich the comparisons among different mechanisms. From Fig. 2,  one also can find that $\tau$ increases obviously as the radius of quantum dot $R_0$ increases from 10 nm to 50 nm. It is highly probable that the carriers' diffusion space is expanded with increasing the radius of the quantum dot, suppressing the occurrence of the Auger-like processes\cite{wx25}. This means that the lifetime $\tau$ can be modulated by the radius of the quantum dot, which is beneficial to adjust the optical properties of quantum dots, such as the luminous intensity and the photoelectric conversion efficiency, and thus improves the performance of devices-based on MHPQDs. We hope these theoretical results could trigger more experimental studies on this aspect.

In conclusion, we theoretically study the ultrafast Auger-like process mediated by three types of quantum defects in MAPbI$_3$ and MAPbBr$_3$ quantum dots. We find that (i)  the lifetime of the whole process maintained in the range of femtosecond is barely sensitive to the trapping depth and the species of defects, so the high tolerance of defects could be well explained by this Auger-like process; (ii) the whole process can be modulated by the radius of the quantum dot.

This work was supported by National Natural Science Foundation of China (Grand No.  11674241).

The data that supports the findings of this study are available within the article [and its supplementary material].

\end{document}